\documentstyle{article}
\begin{document}
\begin{centerline}
{\Large\bf Red quasars not so dusty}
\end{centerline}

\renewcommand{\baselinestretch}{2}
\vspace{3mm}
Benn C.R.$^\ast$, Vigotti M.$^+$, Carballo R.$^\$$, Gonzalez-Serrano, J.I.$^\$$,
S\'{a}nchez S.F.$^\$$

$\ast$  Isaac Newton Group, Apartado 321, 38780 Santa Cruz de La Palma, Spain\\

+ Istituto di Radioastronomia CNR, Via Gobetti 100, I-40100 Bologna, Italy\\

$\$$ Instituto de Fisica de Cantabria (CSIC - Universidad de Cantabria),
Facultad de Ciencias, 
39005 Santander, Spain \\

\vspace{3mm}

\vspace{3mm}

\vspace{3mm}
\rm

\large
{\bf Abstract}

Webster {\it et al\/} (1995)  claimed that up to 80\% of QSOs may be
obscured by dust.  They inferred the presence of this dust from
the remarkably broad range of $B-K$ 
optical-infrared colours of a sample of flat-spectrum PKS radio QSOs.
If such dust is typical of QSOs,
it will have rendered invisible most of those which
would otherwise been have detected by optical surveys.
We used 
the William Herschel Telescope on La Palma to obtain $K$ infrared images of 
54 B3 radio quasars selected at low frequency (mainly steep-spectrum), 
and we find that although several have very red optical-infrared colours,
most of these can be attributed to an excess of light in $K$
rather than a dust-induced deficit in $B$.
We present evidence that some of the infrared excess comes from the light of
stars in the host galaxy
(some, as previously suggested, comes from synchrotron radiation
associated with flat-spectrum radio sources). 
The $B-K$ colours of the B3 QSOs provide no evidence for a large
reddened population.
Either the Webster {\it et al}  QSOs are atypical in having such large
extinctions, {\it or} their reddening is not due to dust;
either way, the broad range of their $B-K$ colours does not provide
evidence that a large fraction of QSOs has been missed from optical surveys.

\section{Introduction}
Internal and line-of-sight dust absorbs some of the light that would
otherwise reach us from extragalactic objects. We can deduce the relative
amounts absorbed
 by comparing the optical colours of objects selected
at a wavelength unaffected by dust absorption (e.g. in the radio).
Recently, Webster {\it et al} (1995) (hereinafter W95)
found a broad range of $B-K$ colours (1 $< B-K <$ 8) for flat-spectrum
QSOs selected from the Parkes radio survey. They interpreted this
scatter in colour in terms of dust-reddening, implying an extinction
in the blue of several magnitudes for a substantial fraction of the
QSOs.  W95 suggested that if this extinction is typical of QSOs,
it will have led to 80\% of
them being missed by optical surveys (and these might contribute that part
of the X-ray background not yet accounted for). This
has profound cosmological implications, e.g. for the space density of
QSOs and its evolution with redshift (Boyle {\it et al} 1988). 

However, several effects might redden the observed $B-K$ colours of
QSOs. Contamination by starlight from the host galaxy,
and by optical/infrared synchrotron radiation associated with flat radio
spectra; variability between the epochs at which the $B$ and $K$
magnitudes were measured; 
and photometric errors,
all need to be excluded before
concluding that dust is responsible for any observed reddening.
In particular, it has long been known that the
optical-infrared colours of  flat-radio-spectrum QSOs can be 
unusually red (Rieke {\it et al} 1979).
The redness is in many cases due to a sharp spectral break in the
optical/infrared, consistent with a high-frequency cutoff in a
synchrotron spectrum (which dominates at radio wavelengths); and is
in many cases not consistent with reddening due to extinction by dust
(Bregman {\it et al} 1981, Rieke {\it et al} 1982).
Serjeant \& Rawlings (1995) have already noted that
such non-thermal emission  may explain the red colours measured by W95.

At low redshift, $z <$ 2, a few red QSOs do exist, and dust is probably 
responsible for the reddening in some cases (Rawlings {\it et al} 1995).
The location of the dust is unknown, 
but it may lie in the obscuring torus popularly
invoked when explaining the diversity of appearance of
active galaxies as a dependence on viewing angle
(Wills {\it et al} 1992, Urry \& Padovani 1995). 
If the torus does not have a sharp edge, some QSOs will be seen 
reddened rather than extinguished.
Some of these reddened objects satisfy the conventional definition of a QSO, 
with broad emission lines in the optical
(e.g. Smith \& Spinrad 1980,
Kollgaard {\it et al} 1995).
Others have only narrow lines in the optical, but have been classified as 
obscured QSOs after
broad $H_\alpha$ emission was discovered in the infrared (e.g. 3C22:
Economou {\it et al} 1995, Rawlings {\it et al} 1995); 
after broad emission lines were detected in polarised light
(e.g. Goodrich {\it et al} 1996, Hines {\it et al} 1995);
or on the basis of X-ray properties (Stocke {\it et al} 1982,
Ohta {\it et al} 1996, Almaini {\it et al} 1995).
Our main interest here is in the cause of reddening of 
objects conventionally classified as QSOs (broad lines in the optical),
since only these would have been included in W95's sample. 

There  exist a number of observations which place limits on the
amount of dust in such QSOs.
One expects for example that significant amounts of dust 
would markedly affect their $U-B$ and $B-V$ colours.
Schmidt (1968) found the variation with redshift of the
$U-B$ and $B-V$ colours of 
(mainly steep-spectrum) 3CR QSOs to be consistent with that expected for
a composite (power-law 
$S_{\nu} \propto \nu^{\alpha}$ plus emission lines) QSO spectrum,
out to z = 2. The scatter on the relations is
$<$ 0.2 mag, implying rest-frame $A_V <$ 0.8 mag.
Smith \& Spinrad (1980) obtained optical spectra for 8 unusually red 3CR QSOs
(15\% of the 3CR QSOs then known) and found 7/8 to have steep straight
spectra whose indices are consistent with reddening $A_V \sim$ 1.5 mag.  
The eighth QSO (representing 2\% of the 3CR QSOs)
has a spectrum which steepens dramatically in the UV, consistent with
considerably higher extinction.
Reddening can also be estimated on the basis of the 
observed ratios of flux in certain emission lines of given elements.
Netzer {\it et al\/} (1995), on the basis of Ly$_{\alpha}$/H$_{\beta}$ 
line ratios, estimated
extinctions up to rest-frame $A_V$ = 1.2 for a sample of 3CR and other
radio-loud QSOs. Larger extinctions have
been inferred on the basis of line ratios for quasars selected from
the Molonglo 408-MHz survey (Baker J.C. \& Hunstead 1995, 1996),
but interpretation of
the ratios is not straightforward (e.g. 
Binette {\it et al} 1993, Baker A.C. {\it et al} 1994)
and considerably less extinction is implied by the small
reddening of the continua; the change in slope from $\alpha$ = -0.5
to $\alpha$ = -1 corresponding to $A_V \approx$ 0.5 mag. 
Boyle \& di Matteo (1995), selecting QSOs in the X-ray rather than
the radio, inferred from the range of optical/X-ray flux ratios a
rest-frame dust extinction $A_V <$ 1 mag. 
These limits are
consistent with the small amounts of dust extinction $\sim$ 0.3 mag 
inferred from the range of optical-UV colours of the optically-selected
PG QSOs
(Tripp {\it et al} 1994, Rowan-Robinson 1995).
Recently, Drinkwater {\it et al} (1996) observed 11 of the redder W95 QSOs at
mm-wavelengths, but failed to detect CO absorption
at a level two orders of magnitude below
that implied if the reddening were due to extinction.  
This suggests that the red colours are
not due to absorption within the galaxy, although they could still be due to
external line-of-sight absorption.
In summary,  a few red QSOs do exist at low redshift, $z <$ 2
(Smith \& Spinrad 1980; Kollgaard {\it et al} 1995), 
but the above results, particularly those of Schmidt, and of Boyle \&
di Matteo, suggest that only a small fraction of QSOs is obscured
by dust with rest-frame extinction $A_V >$ 1 mag, and that
the large dust extinctions deduced by W95 are not
typical of QSOs.
 
Dust obscuration is likely to be more important at high redshift ($z >$ 3)
(Fall \& Pei 1993; Mazzei \& de Zotti 1996; Chini \& Kr\"{u}gel 1994), and 
is probably needed to account for the form of the
faint galaxy counts
(Wang 1991; Gronwall {\it et al} 1995; Campos \& Shanks 1997).
Direct evidence for such dust comes from sub-mm spectra of
high-redshift QSOs
(Omont {\it et al} 1996).

Dust-obscured QSOs have also been considered as a possible
source of that part of the cosmic X-ray background radiation
not already attributed to
known populations (Madau {\it et al} 1994, W95, Comastri {\it et al} 1995).
However, this no longer appears necessary;
after identification of many of the faintest ROSAT X-ray sources
with low-redshift galaxies, the amplitude and spectrum 
of the X-ray background are adequately accounted for
(Boyle {\it et al} 1995,
Carballo {\it et al} 1995,   
Roche {\it et al} 1995, 
Almaini {\it et al\/} 1996, Griffiths {\it et al} 1996).

We have obtained $B-K$ colours and $K$ imaging
of a sample of (mainly steep-spectrum)
B3 QSOs, in order (a) to test the hypothesis that the red colours are
associated mainly 
with flat radio spectra, and (b) to investigate the origin of
any reddening. 

\section{Sample and observations}
The B3
survey 
(Ficarra {\it et al\/}. 1985)
catalogues sources to a radio
flux-density limit $S_{408MHz}$ = 0.1 Jy. 1050 of the sources (the B3VLA
sample, Vigotti {\it et al} 1989) have been
mapped at the VLA in A and C configurations at 1.46 GHz. 
Candidate QSO identifications (objects of any colour, appearing starlike
to the eye)
were sought on the Palomar Observatory Sky Survey
(POSS-I) red plates ($R \leq$ 20).  CCD images were obtained of any 
objects of uncertain classification, or 
falling within 1 mag of the POSS limit, in order to distinguish
reliably between extended and starlike images.  This yielded a sample
of 172 QSOs, the B3-VLA QSO sample, described in detail by Vigotti
{\it et al} (1997).
Optical spectra were obtained for all 172.
125 were confirmed as QSOs, the remainder being stars, galaxies
or BL Lac objects.
The fraction of {\it quasars} fainter than the POSS-I limit depends strongly 
on the frequency of selection and the limiting flux of the sample; at low
frequency this fraction decreases with increasing flux density
(in 3CR, $S_{408MHz} >$ 5 Jy, there are {\it no} QSOs fainter than
POSS-I).
Current optical investigations of the empty fields ($R >$ 20) in the B3 VLA
sample
provide a direct constraint on the number of QSOs missed:
out of 202 POSS-I empty fields with S$_{408} \geq$ 0.8 Jy, 
95 have been optically identified to $R \approx$ 23 - 24 using various
telescopes, and of these, 66 have been  spectroscopically classified as
radio galaxies $0.5 < z < 3.2$, and only 1 as a QSO
(Djorgovski, private communication 1996).
From the histogram of the magnitudes of our 125 confirmed QSOs,
which starts to decline well before
the POSS-I limit, we deduce that {\it at least}
 90\% of QSOs with $S_{408MHz} >$ 0.6 Jy are brighter than the
POSS-I limit (fig. 11 of Vigotti {\it et al} 1989).
We therefore
selected for study all 47 QSOs with  $S_{408MHz} >$ 0.6 Jy, and
7 with 0.5 $< S_{408MHz} <$ 0.6 Jy,
in right ascension range 7 - 14$^h$.

We imaged the sample of 54 B3 QSOs 
in $K$ (2.2 $\mu$m), with the WHIRCAM IR camera of the William
Herschel Telescope on La Palma, on 5 and 6 February 1996
(Carballo et al 1997).
52 of the QSOs were detected; the other two were probably detected,
but unambiguous identification with the radio source was not possible.
The error in measured $K$ is typically 0.1 mag.
The observations, including QSO-subtracted photometry of the host galaxies,
 are reported in detail by Carballo et al (1997).

\section{Results}
The distribution of $B-K$ colours for the B3 QSOs is shown in Fig. 1a
(that for optically-selected QSOs is shown for comparison in Fig. 1b). 
The $B$ magnitudes (accuracy 0.3 mag rms) were taken
from the catalogue of objects on the 
POSS-I plates generated by the Automated Plate-measuring Machine (APM)  in
Cambridge
(Irwin {\it et al\/} 1994). 
The distribution of $B-K$ colours for
B3 QSOs is similar in breadth to that found by W95 for 
flat-spectrum ($S_{\nu} \propto \nu^{\alpha}, \alpha >$ -0.5)
radio-selected QSOs (their fig
1b), except for a lack of extreme red colours $B-K >$ 6. 
As noted in the introduction,  unusually-red optical-infrared colours are often
associated with flat-spectrum QSOs
and are probably due to non-thermal (synchrotron) emission.
There
are only 11 flat-spectrum sources
in the B3 sample.
The distribution of their colours is not significantly different from that
of the steep-spectrum QSOs in the sample, although of the 6 with the flattest
spectra ($\alpha >$ -0.3), only one has $B-K <$ 3.5.

Most of the red B3 QSOs do not have flat radio spectra. We believe
that many appear red because the $K$ light includes a contribution from
stars in the host galaxy. Radio-selected QSOs are usually hosted by
giant elliptical galaxies, and such galaxies have current-epoch 
Mv $\approx$ -23
($H_0$ = 50 kms$^{-1}$Mpc$^{-1}$)
(Lehnert {\it et al} 1992, 
Dunlop {\it et al} 1993).
At the redshifts of the B3 QSOs ($z >$ 0.4), none would
be detectable in $B$ at the POSS limit. 
In $K$ however, as shown in Fig.
2 (the $K-z$ Hubble diagram), 
such galaxies are nearly as bright as many of the QSOs are measured
to be. One therefore expects significant contamination of the $K$
magnitudes by starlight. For example, for the 10 QSOs lying within 1
mag of the mean $K-z$ relation for radio galaxies, the 
starlight from an underlying giant elliptical with $M_V =$ -23 will redden 
$B-K$ by a median
of $\approx$ 1 mag. For $z <$ 1, 
excluding flat-spectrum objects from the
comparison, the mean $B-K$ colour for this group is 3.6, compared with
2.9 for the remaining QSOs, supporting this interpretation.  Further
support for substantial starlight contamination in $K$ comes from the
non-stellar radial profiles of many of the $K$ images of the QSOs,
and from the correlation between  image extension and red $B-K$
evident in Fig. 1.
Image extensions were determined on the basis of deconvolution of 30
images taken on the first night, and
our estimates are conservative; as well as the 14/30 noted in Fig. 1,
several images show weaker evidence
for non-stellar profiles (Carballo {\it et al} 1997).
Images taken on the second night had poorer seeing/focus, and have not been
analysed for extension.

In summary, of the 13 B3 QSOs with measured $B-K >$ 4,
five have extended $K$ images,
two have flat radio spectra ($\alpha >$ -0.5) and one has both.
At
least these eight, and maybe all of the $B-K >$ 4 colours, may thus be
attributed to starlight and/or synchrotron contamination in $K$. 
The sources with $B-K$ = 5.58 (flat-spectrum) and 4.39 (both $K <$ 14)
clearly have a
large excess of $K$ emission relative to other B3 QSOs at similar redshift
and with comparable radio emission.  The red colours of these sources lend
additional support to the hypothesis that the reddening is due to an excess 
of $K$ light rather than a deficit of $B$ light.

Even if a
few of the red colours are not due to starlight or synchrotron contamination, 
other causes must be
explored before attributing the red colours to dust extinction. For example, a
change in the QSO brightness between 1950 ($B$) and 1996 ($K$) would
introduce a random error into the $B-K$ colours.  The size of this
error will typically be a few tenths of a magnitude
(Neugebauer {\it et al} 1979, Meusinger {\it et al} 1994,
di Clemente {\it et al} 1996, Cristiani {\it et al\/} 1996), 
but
variability by 1 - 2 mags on timescales of years is not uncommon
(Elvis {\it et al\/} 1994), 
particularly for low-luminosity QSOs
(Veron \& Hawkins 1995) 
and both extreme blue and extreme red measured B-K colours could
result. Emission lines probably have only a small effect $<\sim$ 0.05
mag on the broad-band colours.  The UV/optical lines in QSOs 
typically
have equivalent widths $\sim$ 50\AA\
(Francis {\it et al} 1991, Miller {\it et al} 1992).
The $H_\alpha$ line might be included in the $K$ band, 
but even with an equivalent width of 1000\AA\ 
(e.g. Hawaii 167, 
Egami {\it et al} 1996), the change in $K$ would only be 0.3 mag.
Only two of the B3 QSOs observed here lie in the affected redshift range
2.05 $< z <$ 2.50; neither has unusual $B-K$ colour.

POSS/APM $R$ apparent magnitudes are also available for the B3 QSOs.
The
$B-R$, $R-K$ colours are
consistent (within the errors, 0.3 mag rms each in $B$, $R$) with power-law
colours -2 $< \alpha <$ 0.  Unfortunately, the $B-R$ and $R-K$ colours  do
not allow discrimination between the effects of reddening and of different
power-law slopes, because at all redshifts of interest, the reddening
vectors are almost parallel to the
locus of power-law colours in $B-R$ and $R-K$.

There is no correlation between $B-K$ colour and redshift, which 
argues against any reddening being due to line-of-sight dust outside the
host galaxy.
This is consistent with the finding of Shaver {\it et al} (1991) that the
observed decrease with redshift of the space density of QSOs is
real and not due to intervening obscuration.

The small incompleteness of the optical identifications for B3 imposes a
slight colour bias.  As noted above, 10\% (5) of the QSOs at this
flux-density level are likely to be fainter than the POSS-I limit
$R$=20, and since there is a weak positive correlation between B-K and
$R$, these are likely to be red. However, there is no reason to suppose
that the colours of these missed objects are due to
effects  other than those discussed above. 

We therefore find no evidence that the redder $B-K$ colours need be
attributed to dust, and the range of observed colours for the
remainder, 2 $< B-K <$ 4, limits the amount of dust reddening in $B-K$ to 
$<$
2 mag, corresponding, for the extinction law of Calzetti {\it et al\/} (1994)
to rest-frame dust extinction $A_V <$ 1.0 mag (z = 2.0) or $A_V <$ 1.6 mag
($z =$ 0.5). This is a conservative limit; one would expect some of the
spread in $B-K$ colours to be intrinsic. This figure is comparable to the limits
on dust
extinction implied by other observations discussed in the introduction. 

\section{Conclusions}
W95 argued that the broad observed range of $B-K$ colours for flat-spectrum
PKS radio QSOs was evidence for dust extinction of up to several magnitudes,
and they argued that if
this extinction is typical of QSOs (i.e. not confined to flat-spectrum
radio QSOs), it implies that a large fraction of QSOs will have been
missed by optical surveys.

We have studied a sample of 54 B3 QSOs,
which  are representative of QSOs with
$S_{408MHz} >$ 0.5 Jy, and we find
a broad range of colours 1 $< B-K <$ 6, similar to that found by W95.
We provide evidence that for both samples of QSOs
the reason
for the range of colours is a variable excess of light in $K$ 
rather than a variable deficit in $B$.
Many of the reddest QSOs in our sample have non-stellar images in $K$,
consistent with underlying giant-elliptical galaxies, and 
indeed most of the B3 and W95 QSOs are not much brighter than one would
expect giant-elliptical galaxies at similar redshift to be (Fig. 2).
This suggests that contamination by 
starlight accounts for much of the reddening of the
measured $B-K$ colours of the QSOs.  
In addition, our data are consistent with some of the
red colours of the W95 QSOs being associated with flat
radio spectra, probably 
due to a steep optical-infrared cutoff in the non-thermal synchrotron spectrum
(Rieke {\it et al} 1979, Rieke {\it et al}
1982, Serjeant \& Rawlings 1995);
we detect no QSOs with $B-K >$ 6, whereas W95 detected several.
Finally, two 
objects which stand out as being particularly luminous in $K$ are also
very red.
We conclude from these results 
that the red colours of at least the B3 QSOs are 
due to
{\it additional} light in $K$ (starlight or synchrotron radiation) rather
than a {\it deficit} in $B$ due to dust extinction.
Either the red colours of the W95 QSOs are due to the same
effects and dust extinction is not important (we incline to this view),
{\it or} the W95 flat-spectrum QSOs are not typical in this respect.
The spread of $B-K$ colours of radio QSOs does not provide evidence
for a large `missing' population of extinguished QSOs. 

\vspace{20mm}
{\large\bf References} 

Aaronson M., 1978, ApJ, 221, L103 \\
Almaini O., Boyle B.J.,  Griffiths R.E, Shanks T., Stewart G.C., 
Georgantopoulos I., 1995, MNRAS, 277, 31\\
Almaini O., Shanks T., Roche N., Griffiths R.E, Boyle B.J., Stewart G.C., 
Georgantopoulos I., 1996, MNRAS, 282, 295\\
Baker J.C., Hunstead R.W., 1995, ApJ, 452, L95 \\
Baker J.C., Hunstead R.W., 1996, ApJ, 468, L131 (erratum) \\
Baker A.C., Carswell R.F., Bailey J.A., Espey B.R., Smith M.G., Ward 
M.J., 1994, MNRAS, 270, 575 \\
Binette L., Wang J. Villar-Martin M., Martin P.G., Magris C.G., 
1993, ApJ, 414, 535 \\
Boyle B.J., Shanks T., Peterson B.A., 1988, MNRAS, 235, 935 \\
Boyle B.J., McMahon R.G., Wilkes B.J., Elvis M., 1995, MNRAS, 272, 462 \\
Boyle B.J., di Matteo T., 1995, MNRAS, 277, L63 \\
Bregman J.N., Lebofsky M.J., Aller M.F., Rieke G.H., Aller H.D., Hodge P.E., 
Glassgold A.E., Huggins P.J., 1981, Nat, 293, 714 \\
Calzetti D., Kinney A.L., Storchi-Bergmann T., 1994, ApJ, 429, 582  \\
Campos A., Shanks T., 1997, MNRAS, submitted  \\
Carballo R., Warwick R.S., Barcons X., Gonz\'alez-Serrano J.I., Barber C.R., 
Mart\'\i nez-Gonz\'alez E., P\'erez-Fournon I., Burgos J., 1995, 
MNRAS, 277, 1312 \\
Carballo R. {\it et al}, 1997, in preparation \\
Chini R., Kr\"{u}gel E., 1994, A\&A, 288, L33 \\
Comastri A., Setti G., Zamorani G., Hasinger G., 1995, A\&A, 296, 1 \\
Cristiani S., Trentini S., La Franca F., Aretxaga I., Andreani P., Vio R., Gemmo 
A, 1996, A\&A, 306, 395 \\
Di Clemente A., Giallongo E., Natali G., Trevese D., Vagnetti F., 1996, ApJ, 
463, 466 \\ 
Drinkwater M.J., Combes F., Wiklind T., 1996, A\&A, 312, 771 \\
Dunlop J.S., Peacock J.A., Savage A., Lilly S.J., Heasley J.N., Simon A.J.B., 
1989, MNRAS, 238, 1171 \\
Dunlop J.S., Taylor G.L., Hughes D.H., Robson E.I., 1993, MNRAS, 264, 455 \\
Eales S.A., Rawlings S., 1996, ApJ, 460, 68 \\
Economou F., Lawrence A., Ward M.J., Blanco P.R., 1995, MNRAS, 272, 5 \\
Egami E., Iwamuro F., Maihara T., Oya S., Cowie L.L., 1996, AJ, 112, 73 \\
Elvis M. {\it et al\/}, 1994, ApJS, 95, 1 \\
Fall S.M., Pei Y.C., 1993, ApJ, 402, 479 \\
Ficarra A., Grueff G., Tomasetti G., 1985, A\&AS, 59, 255\\
Francis P.J., Hewett P.C., Foltz C.B., Chaffee F.H., Weymann R.J., Morris 
S.L., 1991, ApJ, 373, 465 \\
Goodrich R.W., Miller J.S., Martel A., Cohen M.H., Tran H.D., Ogle P.M., 
Vermeulen R.C., 1996, ApJ, 456, L9 \\
Griffiths R.E., Della Ceca R., Georgantopoulos I., Boyle B.J., Stewart G.C., 
Shanks T., Fruscione A., 1996, MNRAS, 281, 71 \\
Gronwall C., Koo D.C., 1995, ApJ, 440, L1 \\
Hewett P.C., Foltz C.B., Chaffee F.H., 1995, AJ, 109, 1498 \\
Hines D.C., Schmidt G.D., Smith P.S., Cutri R.D., Low F.J., 1995, ApJ, 450, L1 
\\
Irwin M., Maddox S., McMahon R.G., 1994, 
  Spectrum (UK Royal Observatories) 2, 14 \\
Kollgaard R.I., Feigelson E. D.,  Laurent-Muehleisen S. A., 
Spinrad H., Dey A., Brinkmann W., 1995, ApJ, 449, 61\\ 
Lehnert M.D., Heckman T.M., Chambers K.C., Miley G.K., 1992, ApJ, 393, 68 \\
Madau P., Ghisellini G., Fabian A.C., 1994, MNRAS, 270, L17 \\
Mazzei P., De Zotti G., 1996, MNRAS, 279, 535 \\
Meusinger H., Klose S., Ziener R., Scholz, R.D., 1994, A\&A, 289, 67 \\
Miller P., Rawlings S., Saunders R., Eales, S.A., 1992, MNRAS, 254, 93 \\
Netzer H., Brotherton, M.S., Wills B.J., Han M., Wills D., Baldwin J.A., 
Ferland G.J., Browne I.W.A., 1995, ApJ, 448, 27 \\ 
Neugebauer G., Oke J.B., Becklin E.E., Mathews K., 1979, ApJ, 230, 79 \\
Omont A., McMahon R.G., Cox P., Kreysa E., Bergeron J., Pajot F., 
Storrie-Lombardi L.J., 1996, A\&A, 315, 1 \\
Ohta K., Yamada T., Nakanishi K., Ogasaka Y., Kii T., Hayashida K., 1996, 
ApJ, 458, L57 \\
Pence W., 1976, ApJ, 203, 39 \\
Rawlings S., Lacy M., Silvia D.S., Eales S.A., 1995, MNRAS, 274, 428\\
Rieke G.H., Lebofsky M.J., Kinman T.D., 1979, ApJ, 232, L151 \\
Rieke G.H., Lebofksy M.J., Wisniewski W.Z., 1982, ApJ, 263, 73 \\
Roche N., Shanks  T., Georgantopulos I., Stewart G.C., Boyle B.J., 
Griffiths R.E., 1995, MNRAS, 273, L15 \\
Rowan-Robinson M., 1995, MNRAS, 272, 737\\
Schmidt M., 1968,  ApJ, 151, 393 \\ 
Serjeant S., Rawlings S., 1995, Nat, 379, 304  \\
Shaver P., Wall J.V., Kellermann K.I., Jackson C.A., Hawkins M.R.S.,
   1996, Nature, 384, 439 \\
Smith H.E., Spinrad H., 1980, ApJ, 236, 419 \\
Stocke J., Liebert J., Maccacaro T., Griffiths R.E., Stiener J.E., 1982, 
ApJ, 252, 69 \\
Tripp T.M., Bechtold J., Green R.F., 1994, ApJ, 433, 533 \\
Urry C.M., Padovani P., 1995, PASP, 107, 813  \\
Veron P., Hawkins M.R.S., 1995, A\&A, 296, 665\\
Vigotti M., Grueff G., Perley R., Clark B.G., Bridle A.H., 
1989, AJ, 98, 419 \\
Vigotti M., Vettolani G., Merighi R., Lahulla J.F., Pedani M., 
1997, A\&AS, 123, 1 \\
Wang B., 1991, ApJ, 383, L37\\ 
Webster R.L., Francis P.J., Peterson B.A., Drinkwater M.J., Masci F.J., 
1995, Nat, 375, 469 \\
Wills B.J., Wills D., Evans N.J., Natta A., Thompson K. L., 
Breger M., Sitko M. L., 1992, ApJ, 400, 96 \\

\vspace{20mm}
{\large\bf Figure captions}

Fig. 1 - Distributions of $B-K$ colours for: 

    (a) Radio-selected B3 QSOs, this paper.  `Extended' indicates that  
the K image is significantly extended.
 `Flat-spectrum' indicates
radio (0.4 - 1.5 GHz) $\alpha >$ -0.5.
Two of the objects marked 
`extended', with $B-K$ = 3.5 and 4.5, are also flat-spectrum.
The two objects for which the K identification is ambiguous have been
omitted from this diagram (and from Fig. 2).
This distribution is consistent with that found for a smaller sample of
(mainly steep-spectrum) Parkes QSOs by Dunlop {\it et al\/} (1989).

    (b) Optically-selected LBQS QSOs, from W95.
This distribution is similar to that found for other optically-selected
samples of
QSOs e.g. that of Elvis {\it et al} (1994).

The difference in the $B-K$ distributions of radio-selected and
optically-selected samples is due mainly to the effects of
apparent-magnitude and colour selection on the latter. The
dashed line superimposed on Fig 1b indicates the colour
distribution of the B3 QSOs after imposing a cutoff $B <$ 19 (similar to
the limiting magnitude of the LBQS sample).
Optical quasar samples are typically
selected on the basis of colour as well, which imposes a further bias.
For example, 
LBQS QSOs, with which W95 compared their data, are selected on
the basis of blue optical colour (Hewett {\it et al} 1995).

(The $B$ passbands used for the samples of Fig 1a and 1b
are slightly different, but
the effect on the measured B-K colour is small, $<$ 0.1 mag for
most of the QSOs.)

Fig. 2 - $K$ apparent magnitude vs redshift for the B3 QSOs. The
number marking the location on the plot of each QSO is its $B-K$ colour. 
Circles indicate B3 QSOs with extended
$K$-band images.  
Underlining indicates B3 QSOs with flat ($\alpha >$ -0.5) radio spectra.
The
curves `B2/6C' and `3CR' show the mean $K-z$ relations
 for B2/6C and 3CR radio galaxies (Eales \& Rawlings 1996).
The former is similar to that for non-evolving giant elliptical
galaxies.
The dashed curve is the mean of those for B2/6C and 3CR;
the dotted line indicates the apparent mag of radio galaxies 2 standard
deviations (1.0 mag) brighter than this mean.
The median 408-MHz flux density of the B3 QSO sample is 2 Jy, close to that
of the B2/6C sample.
The curve `Sbc' shows the locus for typical Sbc spiral galaxies
(spectra from Pence 1976 and Aaronson 1978).
Crosses represent measurements for QSOs from Elvis {\it et al} (1994).

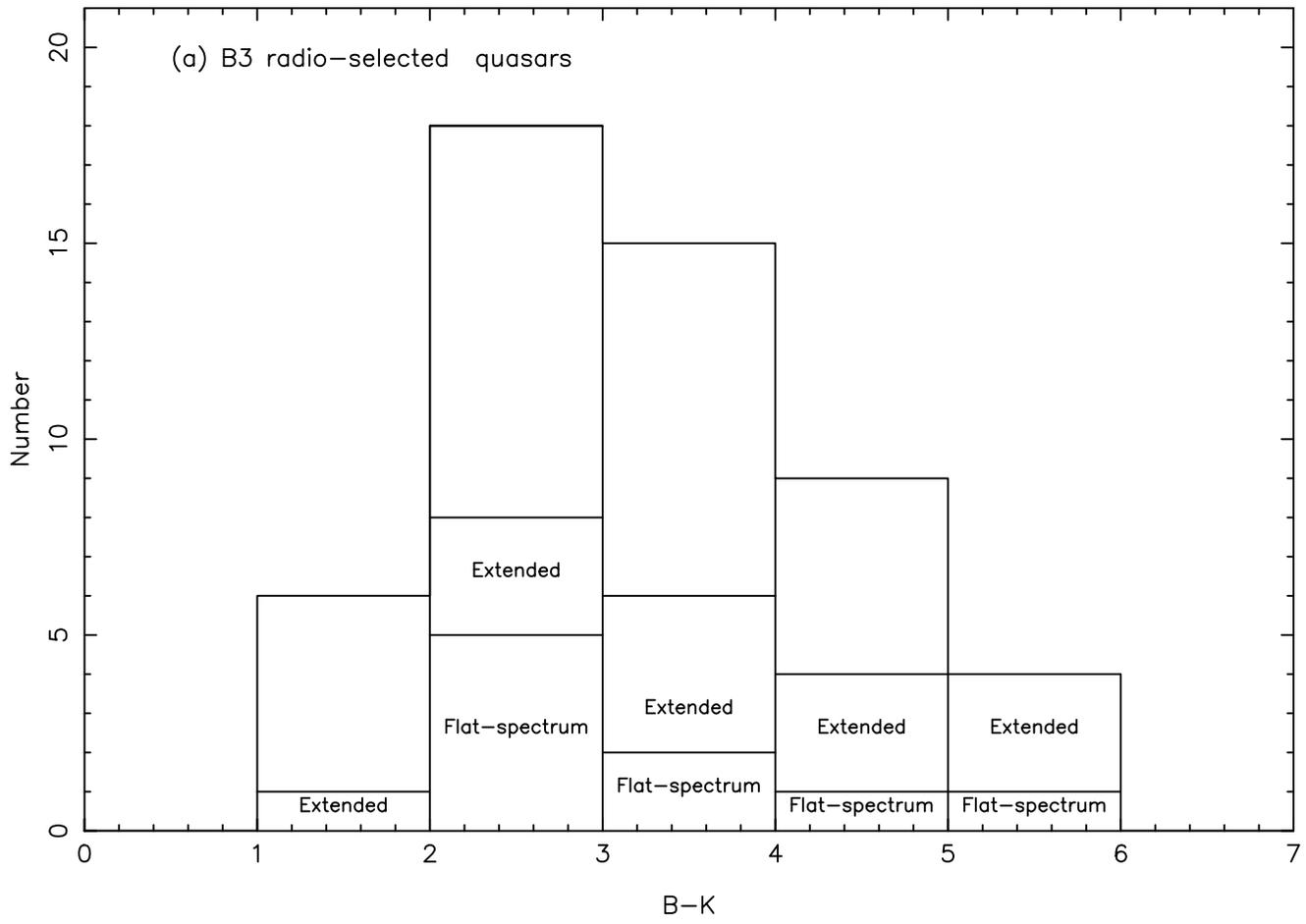
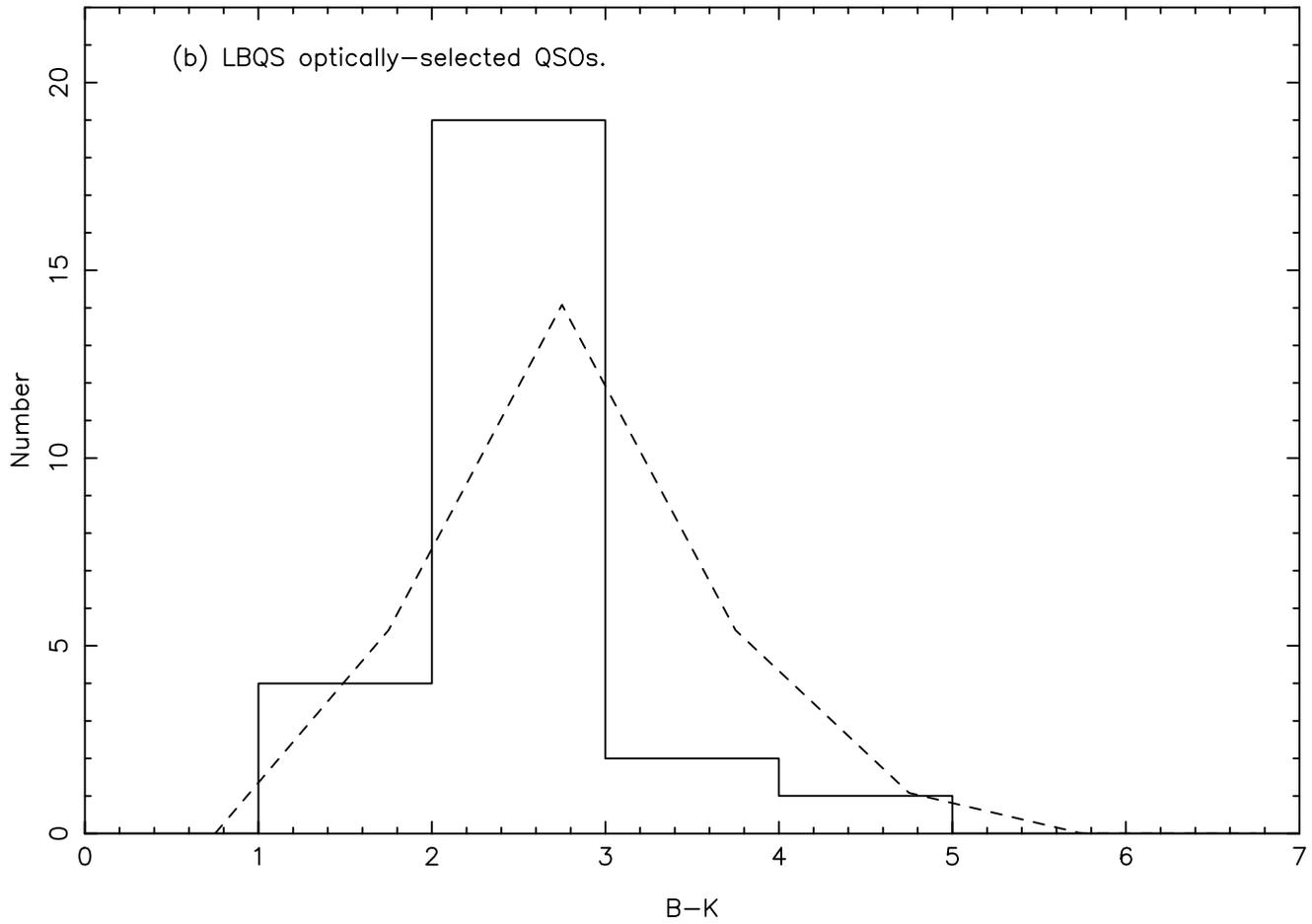
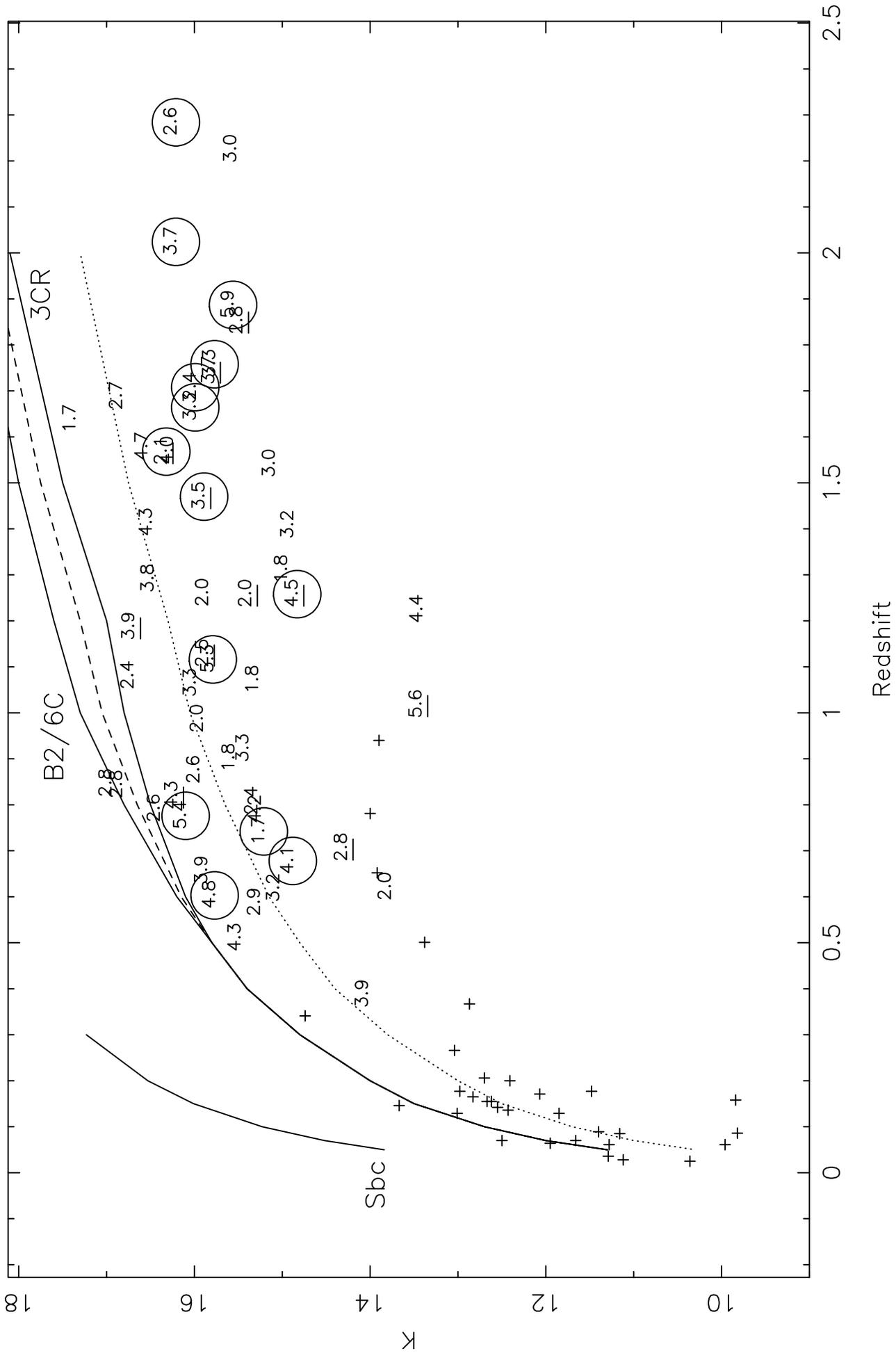
\end{document}